\documentclass[11pt,a4paper]{article}
\pdfoutput=1

\usepackage{jcappub}
\usepackage{multirow}
\usepackage{amssymb} 
\usepackage[normalem]{ulem}

\newcommand{\fett}[1]{\boldsymbol{#1}}

\newcommand{\dd}{{\rm{d}}}
\newcommand{\ii}{{\rm{i}}}

\newcommand{\be}{\begin{equation}}
\newcommand{\ee}{\end{equation}}

\newcommand{\kop}%{{\cal k}}
{\mathfrak{K}}

\definecolor{darkgreen}{rgb}{0,0.6,0}
\newcommand{\Hc}{\mathcal{H}}

\newcommand{\CLASS}{{\sc class}}

\newcommand{\inspire}[1]{\href{http://inspirehep.net/search?p=find+J+#1}
 {{\color{black}[{\color{blue} {\small in}SPIRE}]}}}
\newcommand{\book}[1]{\href{http://inspirehep.net/search?p=#1}
 {{\color{black}[{\color{blue} {\small in}SPIRE}]}}}

\newcommand{\inspired}[1]{\href{http://inspirehep.net/search?p=#1}
 {{\color{black}[{\color{blue} {\small in}SPIRE}]}}}

\newcommand{\HL}{{H_{\rm L}}}
\newcommand{\HT}{{H_{\rm T}}}

\newcommand{\nab}{\nabla}

\makeatletter
\newsavebox\myboxA
\newsavebox\myboxB
\newlength\mylenA

\newcommand*\mybar[2][0.75]{%
    \sbox{\myboxA}{$\m@th#2$}%
    \setbox\myboxB\null% Phantom box
    \ht\myboxB=\ht\myboxA%
    \dp\myboxB=\dp\myboxA%
    \wd\myboxB=#1\wd\myboxA% Scale phantom
    \sbox\myboxB{$\m@th\overline{\copy\myboxB}$}%  Overlined phantom
    \setlength\mylenA{\the\wd\myboxA}%   calc width diff
    \addtolength\mylenA{-\the\wd\myboxB}%
    \ifdim\wd\myboxB<\wd\myboxA%
       \rlap{\hskip 0.5\mylenA\usebox\myboxB}{\usebox\myboxA}%
    \else
        \hskip -0.5\mylenA\rlap{\usebox\myboxA}{\hskip 0.5\mylenA\usebox\myboxB}%
    \fi}
\makeatother

\begin{document}

\title{A Relativistic Interpretation of Bias in Newtonian Simulations}

\date{\today}

\author[a]{Christian Fidler,}
\emailAdd{fidler@physik.rwth-aachen.de}

\author[a]{Nils Sujata and}
\emailAdd{nils.sujata@rwth-aachen.de}

\author[b]{Cornelius Rampf}
\emailAdd{cornelius.rampf@oca.eu}

\affiliation[a]{Institute for Theoretical Particle Physics and Cosmology (TTK), RWTH Aachen University, Otto-Blumenthal-Strasse, D--52057 Aachen, Germany}

\affiliation[b]{Laboratoire Lagrange, UCA, OCA, CNRS, CS 34229, F--06304 Nice Cedex 4, France}

\abstract{
Observables of cosmic structures are usually not the underlying matter field but biased tracers of matter, such as galaxies or halos. We show how the bias found in Newtonian N-body simulations can be interpreted in terms of the weak-field limit of General Relativity (GR). For this we employ standard Newtonian simulations of cold dark matter and incorporate GR/radiation via a weak-field dictionary that we have recently developed. We find that even when a simple local biasing scheme is employed in the Newtonian simulation, the relativistic bias becomes inherently scale-dependent due to the presence of radiation and GR corrections. This scale-dependence could be in principle observed on large scales in upcoming surveys.
As a working example, we apply our methodology to Newtonian simulations for the spherical collapse and recover permille-level agreement between the approaches for extracting the relativistic bias on all considered scales.
}

\maketitle   

\flushbottom
%%%%%%%%%%%%%%%%%%%%%%%%%%%%%%%%%%%%%%%%%%%%%%%%%%%%%%%%%%
\section{Introduction}
\label{sec:intro}

The bias of structures (halos, galaxies) compared to the underlying dark matter distribution 
is a fundamental ingredient for the analysis of large-scale structure surveys, 
and has been extensively studied in both Newtonian \cite{Peacock:2000qk,Scoccimarro:2000gm,Bernardeau:2001qr,Seljak:2004sj,Carbone:2008iz,Matsubara:2011ck,Desjacques:2016bnm}
and relativistic \cite{2009PhRvD..79l3507W,Yoo:2010ni,Baldauf:2011bh,Bruni:2011ta,2011PhRvD..84d3516C} 
frameworks. However, it is not only a requisite to analyse surveys, but 
measuring the luminous galaxies w.r.t.\ the underlying dark matter density offers a 
second observable containing precious information from the large-scale structure. 
This is especially true when observing two (or more) populations that are biased independently \cite{Seljak:2008xr},
and because the bias is sensitive to the primordial non-Gaussianity \cite{Matarrese:2008nc}, 
thereby opening a window into the early Universe. 

On sufficiently small cosmological scales a Newtonian analysis is expected to be accurate, at least in the most simple model of a universe filled only with cold dark matter, 
however the next generation of galaxy surveys will probe large volumes, 
comparable to the cosmological horizon, where a simplified Newtonian analysis 
is no longer accurate. In addition, the early Universe was dominated by radiation, 
including photons and massive neutrinos. The latter are also expected to play an 
important role in structure formation. 
All these contributions are inherently relativistic and thus 
beyond the scope of a Newtonian framework.

During the past years, several approaches to simulate the relativistic evolution of the Universe 
have been developed.
In \cite{Adamek:2013wja,Adamek:2016zes},
employing the relativistic code \texttt{gevolution},
N-body simulations in the weak-field limit of General Relativity (GR) were 
performed (see \cite{Adamek:2017uiq} for the inclusion of massive neutrinos).
Subsequently, photons and massless neutrinos, i.e., purely relativistic species that are of particular importance on large cosmological scales (and rather at early times), 
have been incorporated essentially by three different approaches.
One of these approaches was established by a modified Newtonian N-body code
\cite{Brandbyge:2016raj}, dubbed \texttt{cosira}, 
that makes use of a linear GR dictionary, based on the 
so-called N-body gauge \cite{Fidler:2015npa}.
Another approach was established by a modification of the 
aforementioned relativistic code \texttt{gevolution} \cite{Adamek:2017grt}. 
Lastly, the authors of \cite{Fidler:2016tir,Fidler:2017ebh} introduced 
the Newtonian motion (Nm) gauge approach, originally based on a linear GR dictionary, 
which allows the interpretation of an unmodified Newtonian simulation 
on a perturbed relativistic space-time.
Shortly later, the Nm gauge has been generalised in \cite{Fidler:2017pnb}, 
employing a GR dictionary that is valid within the weak-field limit 
of GR, and furthermore in~\cite{Fidler:2018bkg}  including massive neutrinos.
In the present paper we employ the non-linear Nm approach and study how the bias in GR can be  connected to the bias observed in a Newtonian N-body simulation.

So far, there exists no weak-field description for the bias nor a bias description
that incorporates the effect of radiation --- 
both issues that we aim to address in the present paper.
In \cite{2009PhRvD..79l3507W,Yoo:2010ni,Baldauf:2011bh,Bruni:2011ta,2011PhRvD..84d3516C},
 in the absence of radiation, the relativistic bias has been investigated to linear order in cosmological perturbation theory (CPT) \cite{Kodama:1985bj,Malik:2008im,Villa:2015ppa}.
It has been found that the linear bias receives a GR correction stemming from a long-wave perturbation (itself directly related to the comoving curvature perturbation),
a result that we also recover in the radiation-free and linear limit.
In \cite{Yoo:2014sfa} the relativistic bias has been investigated in a gauge-invariant formulism at second order in CPT, accounting for both volume and source effects.
Recently, some confusion exists in the literature about appropriate gauge choices to 
deal with bias beyond the linear order. In particular, the synchronous-comoving gauge choice (see e.g.\ \cite{Bertacca:2014dra}) 
was discussed in~\cite{Yoo:2014vta} who concluded that this gauge should be avoided for bias definitions,
although, to our opinion, for the incorrect reasons. While~\cite{Yoo:2014vta} claimed that mass conservation is
violated in the synchronous-comoving gauge at second order, it was soon shown in
\cite{Rampf:2014mga,Bertacca:2015mca} that the term that seemingly violates mass conservation is
the consequence of employing a Lagrangian-coordinates approach for the relativistic bias. 
Mass conservation is a fundamental principle in relativistic and Newtonian fluid mechanics and 
holds irrespective of the chosen fluid frame, and therefore we disagree with the 
reasoning of~\cite{Yoo:2014vta}.

Although a second-order bias model in the synchronous-comoving gauge may lead to a consistent description on the largest scales, 
however, for the present purpose of incorporating effects from radiation and the non-linear collapse, this gauge choice should
be avoided, essentially for two reasons. The first reason for that is that the gauge conditions of the synchronous-comoving gauge become inconsistent
when fluids with pressure (e.g., \cite{Villa:2015ppa}; or non-zero vorticity \cite{Rampf:2016wom}) are included.
Secondly, bias relates to the event of matter collapse where non-linearities in the density (and velocity) inevitably become large. 
In a weak-field description this process can be modelled accurately as long as the underlying metric perturbations remain small.
The synchronous-comoving gauge however is not a suitable candidate to 
perform such a weak-field limit as in that gauge the non-linear collapse is 
imprinted in generally large metric perturbations (related to the relativistic Lagrangian displacement field, cf.\ \cite{Rampf:2014mga}).
For similar reasons, also comoving-orthogonal gauges (e.g., the total matter gauge) 
should be avoided for non-linear bias descriptions, since in such gauges the (generally large) velocity 
is proportional to the space-time component of the metric.

Therefore, to include the leading non-linearities in a relativistic bias description, the question of gauge choice becomes important. In the present paper, we discuss certain gauge choices that in particular do not violate the smallness assumptions of the metric perturbations, and additionally allow for large non-linearities in the density and velocity variables. This comes with the advantage that our approach remains accurate 
until much smaller scales, 
but sacrifices some precision on the largest scales (where 
however a linear treatment may suffice). 
Furthermore our approach is not limited to a simple dark matter 
cosmology and allows us to include the impact of photons or neutrinos 
on the relativistic bias.

The paper is organised as follows. In the following we introduce our notations and conventions.
In section~\ref{sec:setup} we provide a summary of the used gauge choices, related to the recently developed Nm framework, as well as briefly discuss the resulting relativistic equations (section~\ref{sec:eom}) and outline our used Eulerian bias description (section~\ref{sec:biasdesc}).
In section~\ref{sec:bias} we discuss the computation of the relativistic bias from Newtonian simulations.
Further results on the bias, in particular the study of the radiation-free (late-time) limit and the case of massive neutrinos, are discussed in section~\ref{sec:furtherbias}.
We confront our theoretical predictions with a relativistic numerical simulation in section~\ref{sec:simulation}, and conclude in~\ref{sec:conclusions}. \\[0.3cm]

\noindent{\bf Notation:} Newtonian quantities, such as the density and velocity (obtained from the Newtonian simulation), are labelled with the index ``N'', whereas relativistic
quantities come with no special index. Biased quantities have the index ``X''.
We set the speed of light to unity.
Most of our equations are formulated in real space, except some of the results in section~\ref{sec:bias} where we occasionally outline results in Fourier space --- noted in the text and through the explicit dependence on the wave vector $|\fett{k}|= k$.

%%%%%%%%%%%%%%%%%%%%%%
\section{Metric convention in our weak-field description}

The metric line element
\be \dd s^2 = g_{\mu\nu} \,\dd x^\mu \dd x^\nu = g_{00}\, \dd \tau^2 + 2 g_{0i}\, \dd x^i \dd \tau + g_{ij} \,\dd x^i \dd x^j
\ee
in a yet unspecified gauge has the following metric coefficients of the scalar type
\begin{subequations}
\label{metric-potentials}
\begin{align}
  g_{00} &= -a^2 \left[ 1 + 2 A \right] \,, \\
  g_{0i} &=  -a^2  \hat\nab_i  B  \,,\\
  g_{ij} &= a^2 \left[ \delta_{ij} \left( 1 + 2 \HL \right) + 2 \left(  \hat\nab_i \hat\nab_j + \frac {\delta_{ij}}  {3} \right) \HT  \right] \,.
\end{align}
\end{subequations}
Here, $a$ is the cosmic scale factor, $\delta_{ij}$ the Kronecker delta,
 we make use of the conformal time $\tau$, and we have defined 
the normalised gradient operator 
$\hat\nab_i \equiv -(-\nabla^2)^{-1/2} \nab_i$ which translates into  $-\ii \hat{k}_i$ in Fourier space, where $\hat k_i \equiv k_i/|\fett{k}|$. 
By employing the normalised gradient operator, we guarantee that all scalar perturbations in the above metric are of the same order in our expansion scheme.

We work in the weak-field limit of general relativity, amounting to a double expansion scheme in metric perturbations and spatial derivatives.
The weak-field approach  encompasses the largest cosmological scales where linear perturbation theory delivers an accurate description, but also incorporates the leading-order contributions from smaller scales where non-linear effects are important. 
We only demand the smallness of the metric perturbations while other perturbations, especially the matter density, may become arbitrarily large. 
See~\cite{Fidler:2017pnb} for a more thorough discussion about our used weak-field definition which is in agreement with the employed weak-field scheme 
of~\cite{Brustein:2011dy,Kopp:2013tqa,Goldberg:2017gsm,Goldberg:2016lcq}.

%%%%%%%%%%%%%%%%%%%%%%%%%%%%%
\section{Relativistic setup and description of bias}\label{sec:setup}

\subsection{Recap of Newtonian motion (Nm) approach}

Let us briefly outline the Nm framework \cite{Fidler:2016tir,Fidler:2017ebh,Fidler:2017pnb,Fidler:2018bkg} which we use as the weak-field dictionary for the Newtonian simulation.
The basic idea of the Nm approach is to enable conventional Newtonian (N-body) 
simulations to obtain a relativistic description that includes the effect of radiation on cosmic structure formation.
Massless radiation can be described  by linear cosmological perturbation theory accurately at all times and scales relevant for structure formation, and consequently the first Nm dictionary employed only a linear dictionary \cite{Fidler:2016tir}. In this approach the linear radiation source provides feedback to the fully non-linear matter field, that itself is evolved within a Newtonian simulation. Subsequently this dictionary was generalised to the weak-field limit. The weak-field Nm framework essentially provides a unified method for resolving the largest scales where linear perturbation theory provides an accurate description as well as the non-linear scales where it includes the leading non-linear contributions.

In the present paper we employ the weak-field Nm formulation, relevant for studying the small scales that lead to the formation of structures. A consistent weak-field
description requires that the  gauge conditions do not violate the underlying smallness assumptions of the metric potentials. For such gauges
the relativistic evolution equations for matter include several relativistic corrections that are not present in a Newtonian description. For example, the relativistic momentum
conservation is coupled to the radiation energy density and to the metric perturbations --- relativistic effects that are not present in the Newtonian momentum conservation (the Euler equation). 

We use the spatial gauge degree of freedom in GR to select so-called Nm gauges, that, by definition, 
imply an equation for the relativistic momentum conservation that precisely matches its Newtonian counterpart.
This way, the GR corrections are embedded in the underlying space-time that can be accounted for as a 
post-processing of the Newtonian simulation simply by using a weak-field Nm dictionary.
For the relativistic equations for matter in the presently employed Nm gauge, see section~\ref{sec:eom}.

\subsection{Used temporal gauge choice}\label{sec:tempgauge}

The Nm idea is related to the spatial gauge fixing, described above. For any temporal gauge choice that does not violate the weak-field smallness conditions, there exists a unique Nm gauge. One possible temporal gauge choice that satisfies this criteria is the Poisson temporal gauge condition (see e.g., \cite{Mukhanov:1990me,Ma:1995ey}) which we will employ in the present paper,
\be
  B = \kop^{-1} \dot H_{\rm T} \,,
\ee
with the operator $\kop = (-\nabla^2)^{1/2}$ which corresponds to the magnitude $k = |\fett{k}|$ in Fourier space. As a consequence of this temporal gauge fixing, the density in our Nm gauge is identical to the density in the Poisson gauge. 

In the remainder of the paper we will discuss different spatial gauges but always keep the temporal gauge fixed by the above gauge condition. This will be advantageous in section~\ref{sec:bias} since the density of tracers does have a non-trivial time-dependence that would complicate temporal gauge transformations. 

\subsection{The N-boisson gauge}\label{sec:Nboisson}

Besides of the Nm gauge we  make use of another gauge in the present paper, the N-boisson gauge. 
We define that gauge through the temporal gauge condition of the Poisson gauge, $B = \kop^{-1} \dot H_{\rm T}$,
while we demand for its spatial gauge that $\HT = 3\zeta$, where $\zeta = \HL + \HT/3 - {\cal H} \kop^{-1} (v-B)$ is 
the gauge-invariant comoving curvature perturbation.
This is a special weak-field gauge in which conventional simulations with CDM and $\Lambda$ can be brought
in agreement with \mbox{GR --- provided} that radiation plays no significant role for the gravitational dynamics anymore. 
For a standard $\Lambda$CDM cosmology, the latter is the case for sufficiently late times ($z < 50$).
In a sense, that gauge behaves as a so-called N-body gauge \cite{Fidler:2015npa}
which implies that for vanishing radiation, the relativistic momentum conservation of matter agrees with the Newtonian Euler equation.
In the spirit of the temporal gauge condition, the N-boisson gauge
has also advantageous properties in the weak-field sense, i.e., all metric perturbations remain naturally small. 

The N-boisson gauge has a different spatial gauge condition as the Poisson gauge, and because of that the analysis in the N-boisson gauge differs slightly from the one of~\cite{Chisari:2011iq} in the Poisson gauge.
Nonetheless, their analysis can be connected to the N-boisson gauge, since in their linear GR dictionary,
after the Newtonian simulation has been terminated, 
N-body particles are displaced to the spatial position of the N-boisson
gauge (see also the related discussions in~\cite{Adamek:2017grt}).

%%%%%%%%%%%%%%%
\subsection{Relativistic equations of motion in the Nm and N-boisson gauge}\label{sec:eom}

As outlined above, we employ the Nm and N-boisson gauges that make use of an identical temporal coordinate, whilst the spatial coordinates are related via a simple spatial gauge transformation $(x^i_{\rm Nb} = x^i_{\rm Nm} + \kop^{-2} \nab^i (H_{\rm T, \, Nm} - 3\zeta))$. 

To outline the difference between these two gauges it is instructive to 
fix for the moment only the temporal gauge condition. In such a not yet completely fixed gauge the relativistic momentum conservation of CDM reads 
\begin{align} \label{eq:relatnoSF}
  \left[ \partial_\tau  + v^j_{\rm cdm}  \nab_j \right] v_i^{\rm cdm} &= - \Hc v_i^{\rm cdm}  - \nab_i A  %\nonumber \\ &
+  \frac{2}{3\rho} \nab_i \Sigma^{\rm cdm}- \left(\partial_\tau +\Hc\right) \kop^{-2} \nab_i \dot{H}_{\rm T} 
\intertext{(vector perturbations are suppressed for simplicity),
where $v_i =  \nab_i v^{\rm cdm}$, and $\Sigma^{\rm cdm}$ is the anisotropic stress resulting from overlapping matter streams.
The above equation is simply relating the gravitational acceleration of fluid trajectories (the l.h.s.) 
to the Hubble drag term, the anisotropic stress 
and certain metric perturbations (all r.h.s.). 
This relativistic result (derived in~\cite{Fidler:2017pnb}) should be contrasted with
the Newtonian momentum conservation which reads}
  \left[ \partial_\tau  + v^{j}_{\rm cdm,N}  \nab_j \right] v_i^{\rm cdm,N} &= - \Hc v_i^{\rm cdm,N}  - \nab_i  \Phi^{\rm N}  +  \frac{2}{3\rho} \nab_i  \Sigma^{\rm cdm,N} \,,
\end{align}
where the index ``N'' denotes Newtonian quantities (as determined through Newtonian simulations), and $\Phi^{\rm N}$ is the cosmological potential subject to the usual Newtonian Poisson equation.

At this stage, having eq.\,\eqref{eq:relatnoSF} as a starting point gives us two options that allows us to find relativistic trajectories that are formally identical with the Newtonian trajectories:
\begin{enumerate}
 
  \item We fix the spatial gauge by demanding that the relativistic corrections on the r.h.s.\ of~\eqref{eq:relatnoSF} are exactly identical with the Newtonian potential, i.e.,  
   \be
      A + ( \partial_\tau + \Hc) \kop^{-2} \dot H_{\rm T} = - \Phi^{\rm N}\,, \quad \hspace{2cm} \text{(spatial Nm gauge condition)} 
   \ee
  leading to the relativistic momentum conservation in the Nm gauge which is formally identical with the Newtonian momentum conservation, {\it even at times when radiation can not be neglected}.
 
 \item We fix the spatial gauge according to 
   \be
     \HT = 3\zeta\,,  \quad \hspace{4.3cm}\text{(spatial N-boisson gauge condition)} 
   \ee
   completing the gauge fixing for the N-boisson gauge, which also results in having a Poisson equation in Newtonian form (see below). 
   The resulting relativistic momentum conservation then still contains 
   a relativistic term $\sim \left(\partial_\tau +\Hc\right) \kop^{-1}\hat\nab_i \dot \zeta$, however, crucially, {\it that relativistic term vanishes when the effect of radiation can be neglected}. This is precisely the case at sufficiently late times ($z < 50$), which thus makes the N-boisson gauge a suitable gauge for analyses at late times.

\end{enumerate}
As regards to the mass conservation, since the temporal gauge conditions in the discussed two gauges are identical, mass conservation for both the Nm and N-boisson gauge reads
\be
  \partial_\tau \delta^{\rm cdm}_{\rm count} +  \nabla^i \cdot \left( \left[ 1+ \delta^{\rm cdm}_{\rm count} \right] v_{i}^{\rm cdm} \right) = 0 \,,
\ee 
where we have defined a new density, {\it the counting density},
\be  \label{eq:counting}
 \delta^{\rm cdm}_{\rm count} = \delta + 3 \HL\,,
\ee
with $\delta = (\rho -\bar\rho)/\bar \rho$ being the actual relativistic density contrast. 
The counting density will be useful later when relating relativistic densities to the density as counted in Newtonian simulations.

Finally, note that the Nm spatial gauge condition features a residual gauge degree of freedom, which suitably can be linked to the initial conditions (ICs) that are used to initialise the Newtonian simulation; see section 4.1.\ in~\cite{Fidler:2016tir} for a thorough discussion on ICs.
In the present paper, we fix the remaining gauge freedom by identifying the spatial gauge at the initial time with the N-boisson gauge.

%%%%%%%%%%%%%%%%
\subsection{Eulerian bias description}\label{sec:biasdesc}

In the present paper we investigate a dictionary that links results
from Newtonian matter simulations to a relativistic bias description.
That dictionary also requires the specification of a Newtonian bias scheme that relates the simulated Newtonian density to the tracer density obtained in that simulation.
We employ a local Eulerian bias scheme, defined in real space via \cite{Desjacques:2016bnm}
\be
 \delta_{\rm X}^{\rm N} = b_1 \delta_{\rm N} + \frac 1 2 b_2 \delta_{\rm N}^2 + \frac 1 2 b_s s_{\rm N}^2 \,,  
\ee
where $\delta_{\rm X}^{\rm N}$ is the density of the tracer density X (e.g., halos, galaxies), $b_1$, $b_2$ and $b_s$ are constant Newtonian bias coefficients, $\delta_{\rm N}$ the fully non-linear density as measured from Newtonian simulations, and
$s_{\rm N} = 2 \partial_i \partial_j \Phi_{\rm N}/(3\Hc^2) - \delta_{ij} \delta_{\rm N}/3$
a function that extracts the tidal effects from the non-linear density.
Here, $\Hc$ is the conformal Hubble parameter, $\Phi_{\rm N}$ the Newtonian cosmological potential satisfying the usual Poisson equation, and $\partial_i$ the Newtonian spatial derivative w.r.t.\ Eulerian component $x_i$.

%%%%%%%%%%%%%%%%%%
\section{Bias: from General Relativity to Newtonian Simulations and return} \label{sec:bias}

Having introduced the necessary ingredients, we are now ready to address the main topic of the present paper, i.e., how to link the Newtonian bias to General Relativity. In particular, we seek to find the relativistic counterpart of $\delta_{\rm X}^{\rm N}$, the relativistic tracer density that we call $\delta_{\rm X}$ in the following.

First, we can embed a Newtonian simulation into General Relativity by using the Nm gauge. This embedding is uniquely defined by the ICs of the corresponding Newtonian simulation. This means that the measured bias parameters from a Newtonian simulation are linked to exactly one Newtonian motion gauge.
Employing the weak-field dictionary the relation between the Newtonian density, $\delta^{\rm N}$, and the relativistic density, $\delta$, 
is
\be \label{eq:dict}
\delta^{\rm N} = \delta + 3\HL\,,  %\quad \quad \quad \Phi^N = \Phi + \gamma
   \qquad \hspace{3cm} \text{(Nm~and~N-boisson~gauge, with~radiation)}
\ee
which evidently agrees with the aforementioned counting density defined in eq.\,\eqref{eq:counting}.

While the pure matter densities are related to their GR counterparts by this very simple dictionary, a similar relation for the tracer density $\delta_{\rm X}$ does not exist in general. This question is linked to understanding the process of tracer formation from the underlying dark matter distribution including the impact of relativistic corrections and may depend significantly on the chosen gauge.

In our case of considering purely spatial gauge transformations, using a Newtonian motion gauge leads to small deformations of the coordinate system beyond the homogenous expanding space-time assumed in the Newtonian simulations. While this is an important correction on large scales close to the horizon, we find that in all
Nm gauges that are weak-field compatible (cf.\ the discussion in~\cite{Fidler:2017pnb}), 
the physics on non-linear scales is not affected by relativistic corrections. For example while the distance between two distant objects in GR may be significantly different from the distance found in a Newtonian simulation, local distance measurements remain almost unaffected.  

From this it follows that a halo finder that is taking into account all the information present in GR will find exactly the same halos as the Newtonian one. The reason being that the process of defining a halo is completely local and the relativistic interpretation may change the distances between halos, but not the local physics. This statement remains true as long as $\HT$ is small compared to the non-linear matter densities on scales that are relevant for the formation of the tracers.

We thus can employ the simple geometrical relation, taking into account the 
volume deformation (i.e., the deformation of the spatial 3-volume due to relativistic diagonal corrections) that alters the mass density according to 
\be \label{eq:Halo}
\delta_{\rm X}^{\rm N} = \delta_{\rm X} + 3\HL \,. \hspace{3.3cm} \text{(Nm~and~N-boisson~gauge, with~radiation)}
\ee
The last equation is one of our main results in the present paper; it provides the 
link between the Newtonian and relativistic tracer density.
This equation states that the tracer density is only changed by the large-scale GR corrections that move around individual halos compared to their Newtonian positions, but the GR dictionary does not change whether a local structure is identified as a halo/galaxy. 

Since we consider a  local Newtonian bias scheme, which amounts at the leading order to $\delta_{\rm X}^{\rm N} = b_1 \delta^{\rm N}$, we find the following simple relation for the relativistic bias
\be \label{eq:deltaX-Nm}
\delta_{\rm X} = b_1 \delta + 3\HL (b_1 - 1) \,, \hspace{1.8cm} \text{(Nm~and~N-boisson~gauge, with~radiation)}
\ee
where we have used our GR-dictionary~\eqref{eq:dict} and the above Eq.~\eqref{eq:Halo}.
Observe that
an additional bias proportional to the local volume perturbation $\HL$ appears on top of the Newtonian bias $b_1$ that is now multiplied with the relativistic density. 
This additional term describes that in GR, not only high-density regions can collapse, but also regions that start on a large volume perturbation but with a somewhat smaller initial matter overdensity. This is directly opposed to the Newtonian case where a collapse is uniquely related to the initial overdensity alone. This additional correction from GR to the relativistic bias is multiplied by $(b_1 -1)$ stating that it results from a source that does not follow the dark matter particles or halos; it naturally has to vanish for tracers with bias $b_1 = 1$.  
%%%%

A key signature of the relativistic bias is its generic scale dependence.
Transforming for the next few equations into Fourier space, we can recast the 
GR bias into a bias based on the relativistic density $\delta$ alone:
\be
\delta_{\rm X}(k) = b_1 \delta + 3\HL (b_1 - 1) \simeq b_{\rm GR}(k) \delta(k) \,,
\ee
with
\be \label{eq:bias-exp}
b_{\rm GR}(k) =  b_1 + 3 (b_1 -1) \frac{T_{\HL}}{T_{\delta}}(k) \,,
\ee
where  $T_{\HL}$ and $T_{\delta}$ are respectively the transfer functions of $\HL$ and $\delta$.
The relativistic bias is necessarily scale dependent due to the different evolution of the density and volume perturbation. 

Including the higher-order terms for the relativistic bias is straightforward. Using the GR dictionary and the same steps as outlined above, we find in real space 
\be \label{delta2}
\delta_{\rm X} = b_1 \delta + 3\HL (b_1 - 1) + \frac 1 2 b_2 (\delta + 3\HL)^2 + \frac 1 2 b_s \left(\frac {2}{3\Hc^2} \partial_i \partial_j (\Phi +\gamma) - \frac 1 3 \delta_{ij} (\delta+ \HL)\right)^2  \,,
\ee
with 
$\gamma \equiv - (\partial_\tau + \Hc)\kop^{-2}\dot{H}_{\rm T} + 8\pi G a^2 \kop^{-2}\Sigma$ and the Bardeen potential $\Phi$, where it should be noted that, 
due to the used spatial gauge fixing, in the last term $\Phi +\gamma$ is nothing but $\Phi_{\rm N}$.
Our analysis is consistent to leading order with weak-field relativity, 
however some of the terms in~\eqref{delta2} are subdominant and may be neglected.
We find that $\HL$ is weak-field suppressed compared to $\delta$ while $\gamma$ is suppressed relative to $\Phi$. This implies that we may neglect all relativistic corrections in the higher-order bias. 
We thus have to a good approximation
\be \label{approx:b2}
\delta_{\rm X} \simeq b_1 \delta + 3\HL (b_1 - 1) + \frac 1 2 b_2 \delta^2 + \frac 1 2 b_s s^2  \,,
\ee
with $s_{ij} = (2\partial_i \partial_j \Phi/(3\Hc^2) -  \delta_{ij} \delta/3)$. 
The same necessarily applies not only to second order, but to all higher-order corrections in the weak-field limit.

It should be noted that the bias in general relativity does depend on the chosen gauge and this is the bias in the Newtonian motion gauge corresponding to the simulation that was used to determine $b_1$, $b_2$ and $b_s$. However, the Newtonian motion gauge shares the temporal gauge choice of the Poisson gauge and therefore this relation may be connected to the Poisson gauge bias with very little effort. The densities and the potential $\Phi$ remain identical, while only $\HL$ has to be replaced with $\Phi -  H_{\rm T}/3$.

%%%%%%%%%%%%%%%%%%%%%%
\section{Further results on the relativistic bias}\label{sec:furtherbias}

Most studies of bias in general relativity employ a simplified Universe, dominated by matter plus a cosmological constant. For a standard $\Lambda$CDM Universe, this simplification is valid at sufficiently late times.
This makes for a particularly interesting and simple special case that we discuss in
the following section. Then, in section~\ref{sec:photons} we switch on radiation and provide the full analysis.

\subsection{Bias in the radiation-free (late-time) limit}

In the case of vanishing radiation there exists a remarkably simple Newtonian motion gauge, i.e., the N-boisson gauge (see section~\ref{sec:Nboisson}) with $\HT = 3\zeta$ and $\zeta$ the comoving curvature perturbation, for which the densities are related via
\be  
  \delta^{\rm N} = \delta + 3 (\Phi - \zeta)\,,  \qquad \hspace{4cm} \text{(N-boisson~gauge,~no~radiation)}
\ee 
where $\Phi$ and $\zeta$ can be easily obtained by using standard linear Boltzmann codes. By contrast, to derive the density in the case where radiation can not be neglected, one is forced to calculate $\HT$ which is computationally more demanding (cf.\ \cite{Fidler:2017pnb}).

Traditionally, the relativistic bias has been investigated in the relativistic counterpart of a Lagrangian-coordinates approach \cite{2009PhRvD..79l3507W,Bruni:2011ta}. For vanishing vorticity (cf.\ \cite{Rampf:2016wom}) that Lagrangian approach is established in the synchronous-comoving gauge. To linear order the relativistic Lagrangian bias is
\be \label{eq:synch}
\delta_{\rm X} = b_1^{\rm L} \delta \,,  \qquad \quad \hspace{1cm} \text{(synchronous-comoving~gauge,~no~radiation,~linear~order)} 
\ee
where $b_1^{\rm L}$ is the relativistic bias in Lagrangian coordinates.

Now we show how this standard linear equation can be related to our findings (which however hold beyond linear theory). For this we employ the Nm dictionary 
to connect the simulation density to the N-boisson gauge density,
which, as discussed above, is identical with the density in the Poisson gauge
\be\label{eq:Poisson-bias}
 \delta_{\rm X} = b_1 \delta + 3(\Phi - \zeta) (b_1 - 1)  \,.  \qquad \quad \text{(Poisson~and~N-boisson~gauge, no radiation)} 
\ee 
There are two simple but important observations to make the connection of the last result to~eq.\,\eqref{eq:synch}. First, observe that the $3(\Phi - \zeta) = \Hc \kop^{-1} v$ is nothing but the temporal gauge generator that connects the Poisson (or N-boisson) gauge to the synchronous-comoving gauge.
Secondly, from Newtonian bias arguments \cite{Matsubara:2011ck} it is well known that the linear Eulerian bias is related to the linear Lagrangian bias via 
\be
  b_1 -1 = b_1^{\rm L} \,,
\ee
a relation that makes much sense also in the present case. Indeed, the quantity $3(\Phi - \zeta)$ physically amounts to the {\it local} volume deformation, a {\it Lagrangian quantity;} explaining why the bias factor attached to $3(\Phi - \zeta)$ is a Lagrangian one.
Furthermore, this directly shows that the two relations are compatible and that the simulation density $\delta^{\rm N} = \delta + 3 (\Phi - \zeta)$ is in fact identical to the synchronous-comoving gauge density (again, vanishing radiation and linear treatment is assumed).

The spatial coordinates of the Nm gauge can be understood as representing a Fermi frame~\cite{Baldauf:2011bh,Desjacques:2016bnm} in which the local evolution is now Newtonian on the small scales. The evolving metric of the Newtonian motion gauge on the other hand traces the divergence between separated Fermi frames and patches them back into a global space-time. 
Consequently the bias found in a Newtonian simulation is the scale-independent bias of that particular local Fermi frame and our analysis allows us to construct a relativistic bias from these.   

We have thus established a simple relation how the linear relativistic bias can be interpreted in other gauges.
Before considering the case with non-vanishing radiation in the following section, let us briefly comment on the interpretation beyond the linear approximation. 
There has been significant work on bias to second order in perturbation theory (see \cite{Yoo:2014vta,Bertacca:2015mca}), but we work to leading order in weak-field gravity, where densities and velocities may become large while the metric potentials remain small. Our work thus is valid up to the very small scales, while our approach does not include second-order metric corrections on the large scales. 

Within the weak-field limit, the synchronous-comoving gauge is no longer a viable choice, since the metric perturbations do not remain small and in fact are responsible for the non-linear the growth of structures. The Nm gauge (or the Poisson gauge), on the other hand, remains well defined and can be used to describe the bias consistently to weak-field precision. For this reason our formulation describes the physics accurately on all relevant scales including non-linearities, while a description in synchronous-comoving gauge should be avoided on the non-linear scales due to its incompatibility with the weak-field approach.

\subsection{The impact of relativistic species on bias}\label{sec:photons}

As evident by a host of astrophysical observations,
the early Universe is radiation dominated, and, depending on the initial redshift considered in the simulation, radiation might introduce a non-negligible correction.
In this case we can no longer use the N-boisson gauge, but have to use a dynamically evolving Nm gauge, leading to the bias (obtained from replacing $\zeta \to \HT$ in~\eqref{eq:Poisson-bias}):
\be\label{eq:bias-NM}
\delta_{\rm X} = b_1 \delta + (b_1 - 1) (3 \Phi - \HT) \,,  \qquad \hspace{3cm} \text{(Nm~gauge,~with~radiation)}
\ee
where $\HT$ is related to the spatial gauge transformation connecting the Nm gauge to the Poisson gauge.
The simple relation to the synchronous gauge that we had found in the pure matter case is lost in the presence of radiation. The reason for this is because the synchronous-comoving gauge is no longer comoving to the matter particles in the presence of additional species that possess a pressure. Calculations of the bias based on the synchronous-comving gauge therefore can no longer be applied in more complex cosmologies, while our method (which is closely related to the Poisson gauge) can be adapted. 

In the case of massive neutrinos, we found in a recent paper \cite{Fidler:2018bkg} that neutrinos have a non-trivial impact on structure formation on the smaller scales (also within the Nm gauge framework).
The above assumption \ref{eq:Halo} for the bias can thus no longer be used since they neglect the non-trivial impact neutrinos have on the small-scale structure formation (especially for large neutrino masses). A solution to this is obtained by using the specifically designed backwards approach presented in~\cite{Fidler:2018bkg}. In the following we briefly summarise the ingredients for that approach, and refer to the aforementioned papers for further details.

The so-called backscaling approach to N-body simulations makes use of ICs at time $z=0$ which are then appropriately rescaled by the linear-growth function to the desired initialisation redshift. This way, not the actual ICs of the universe are prescribed, but instead of a fictitious universe that has the same radiation content as the Universe at redshift $z=0$.
This backscaling approach is actually the standard to initialise N-body simulations, since these simulations generally do not evolve for radiation perturbations. In~\cite{Fidler:2017ebh} it has been shown, that N-body simulations from such backscaled ICs can be self-consistently embedded in a relativistic space-time.

For massive neutrinos the situation is more complex and in~\cite{Fidler:2018bkg} a backwards approach for neutrinos was described, that coincides with backscaling only in the limit of massless neutrinos. It allows the absorption of small scale effects directly into the ICs and thus treat them at the level of the Newtonian simulation and not as a modification of the space-time. 
This in turn can be used to solve the problem mentioned above. By employing a backwards method, the complex effects of neutrinos are automatically included in the Newtonian simulation and already present in the 'Newtonian' bias obtained from such a simulation. We are then able to relate the bias from such simulations to the relativistic Poisson-gauge bias using the simple relation~\eqref{eq:Poisson-bias}.

\section{Numerical Simulations} \label{sec:simulation}

Our analysis connects the non-linear bias found in Newtonian simulations to the relativistic weak-field bias in Poisson gauge. However it also holds for other types of Newtonian simulations. In order to show the impact of relativistic corrections, we perform a suite of simulations for the spherical collapse of a top-hat overdensity, both using Newtonian equations and using general relativity in the Poisson gauge. 

We consider an isolated spherical mass shell characterised by a small-scale overdensity $\delta_S$ in a background Universe perturbed by a long mode $\delta_L$. By defining the mass with respect to the shell's physical volume $M_S = \int_{V_R} \dd^3R \, \delta\rho_S $ we receive the shell mass, which sources our local gravitational potential. The geodesic equation for the physical radius of the shell $R$ reads in the weak-field limit
\be 
\frac{\ddot R}{R} - \frac{\ddot a}{a} - \ddot \Phi - 2 \frac{\dot a}{a} \dot \Phi  + \frac{M_S G}{R^3} = 0 \,,
\label{eq:Gr-collapse}
\ee
where an overdot denotes a partial derivative w.r.t.\ conformal time.
The last term describes the shell mass driving the collapse, while the Hubble drag $\frac{\ddot{a}}{a}$ describes the impact of the background evolution on the shell. 
The contributions $\ddot{\Phi} + 2 \frac{\dot{a}}{a} \dot{\Phi}$ is the relativistic acceleration of the shell including the impact of the long mode in GR. For our simulations in Newtonian gravity we neglect these terms, while we keep them otherwise. 

To compute the bias we fix the initial time $z_{ini} = 100$ and the time of collapse $z_{coll} = 3$, and solve the equation for a structure mass of $M = 10^{13} M_{sun}$ for varying ICs $\delta_S^{ini} \in [0.14, 0.15]$ and $\delta_L^{ini} \in [0.01, 0.02]$. Each simulation corresponds to an initial over-density of a radius $R$ embedded in a background consisting of a long mode with an evolution precomputed in \CLASS~\cite{Blas:2011rf}. We then solve the differential equation numerically to obtain the evolution of the radius $R$. The chosen initial values correspond to different collapse times and we can extract the required relation between the small and large mode that will provide a collapse at $z_{coll} = 3$. In combination with a halo mass function this can finally be converted into the resulting bias. Note that for the relative comparisons that we perform here, the mass function is not relevant as long as it is chosen identically in the relativistic and Newtonian case. It is however important for the absolute value of the bias. For more details about the method see \cite{LoVerde:2014pxa}. 

\begin{figure}[htbp]
	\centering
	\includegraphics[width=0.95\textwidth]{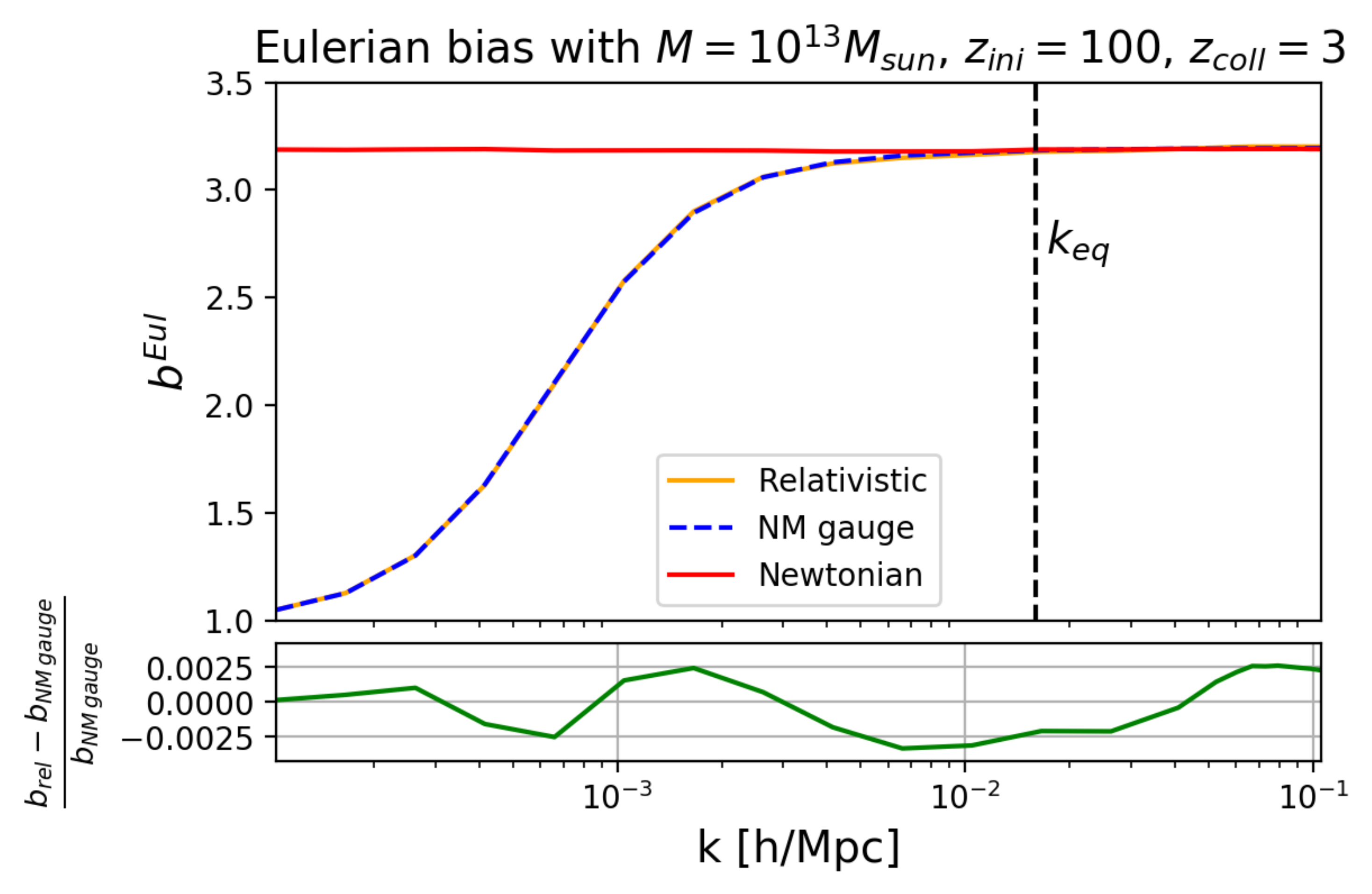}
	\caption{The top plot displays the different methods of determining the halo bias. The orange and the red line are given by the relativistic and Newtonian simulation respectively, the black dashed line depicts the matter radiation equality scale. The blue dashed plot showcases the transformation from Newtonian to relativistic bias using Eq.~\eqref{eq:bias-NM} in the Newtonian motion gauge framework. The relative precision of both approaches is illustrated in the bottom green plot.}
	\label{fig:NMG_bias}
\end{figure}

\begin{figure}[htbp!]
	\centering
	\includegraphics[width=0.95\textwidth]{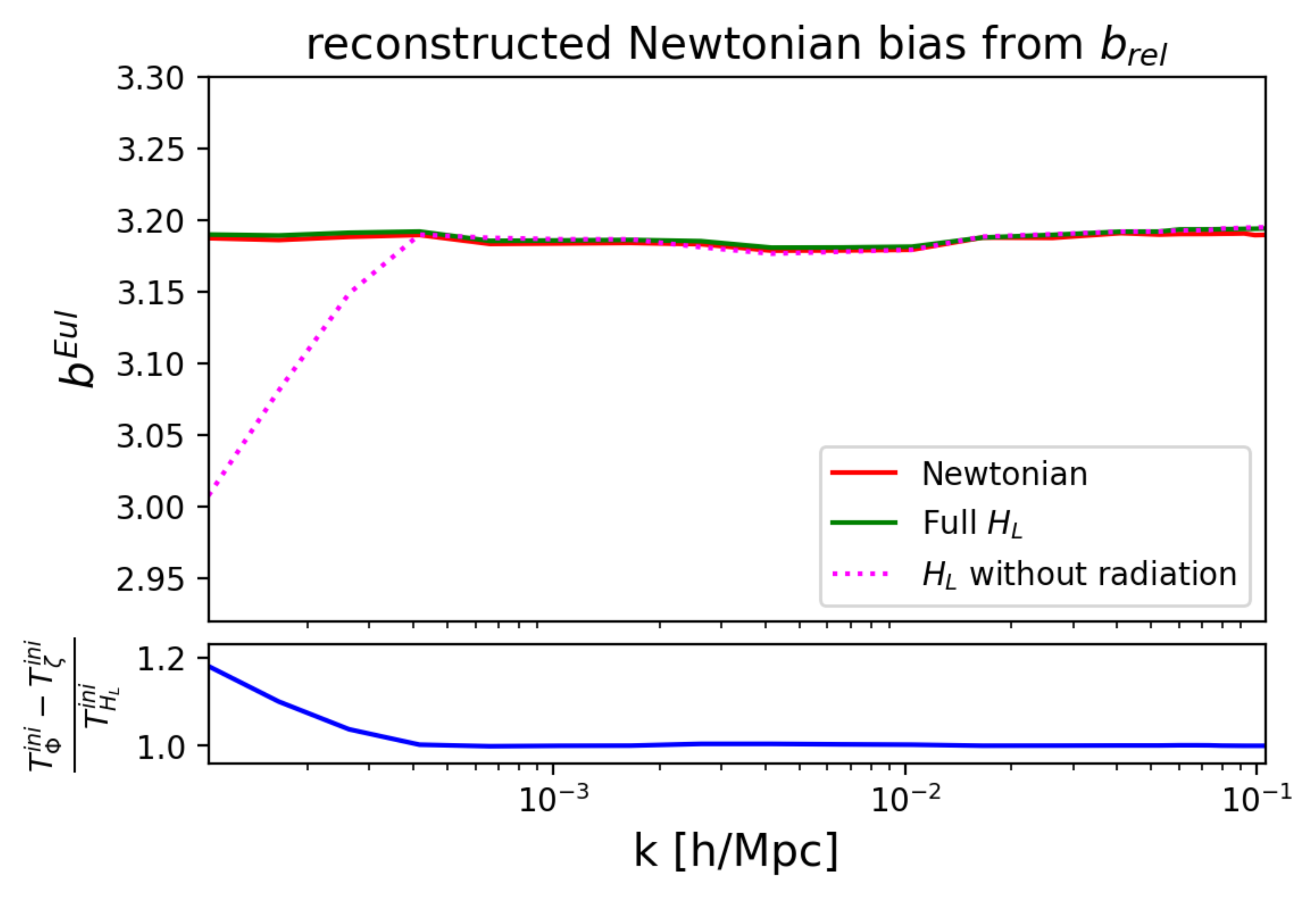}
	\caption{Shown is the reconstruction of the Newtonian bias starting from the relativistic one. In the top plot the calculation using the full $\HL$ is displayed in green while the magenta dotted line depicts the pure matter computation $\HL = \Phi - \zeta$. The bottom plot illustrates the ratio between the transfer functions of the two cases at $z_{ini}$. Note that the axes of this plot span a much smaller range than in the previous one. The Newtonian bias is scale independent within the numerical error of our simulation. }
	\label{fig:NMG_bias_backwards}
\end{figure}

Our results are shown in figure~\ref{fig:NMG_bias}. We find that the Newtonian case leads to a nearly scale-independent bias as expected. Employing the full Poisson gauge dynamics however leads to a significant modification of the bias on scales larger than the matter-radiation equality scale. We then compare this with the relativistic bias obtained from interpreting the Newtonian one using Eq.~\eqref{eq:bias-NM}. The Newtonian motion gauge interpretation of the Newtonian bias agrees with the relativistic Poisson gauge bias at the permille level on all analysed scales. The predicted scale dependence of the relativistic bias is clearly visible and is matched by the relativistic simulation perfectly. 

We further may investigate the impact of relativistic species on the bias using our simulations. Instead of the N-boisson gauge that should be used in a pure matter Universe, we may employ the dynamical Newtonian motion gauge corresponding to the initialisation time of $z=100$ in Eq.~\eqref{eq:bias-NM}. We find that the effect is negligible on the relativistic bias; at least for initial times of redshift $100$. But we may still confirm that our method does compute radiation effects consistently. Since the large-scale relativistic bias becomes very close to unity, small corrections can no longer be seen. However we may also reconstruct the Newtonian bias from the relativistic one using the inversion of Eq.~\eqref{eq:bias-exp}
\be
b_{\rm Newtonian}(k)  =  \frac{b_{\rm GR} +  T_\HL/T_\delta}{ 1 + 3 T_\HL/T_\delta }  \,,
\ee
that returns an almost scale-invariant result that can be compared against the Newtonian simulation.
The outcome is illustrated in figure~\ref{fig:NMG_bias_backwards}.
On the small scales both methods are consistent, but on the large scales we only find a perfect agreement between the reconstructed Newtonian bias and the numerical simulation when including also the impact of radiation on the Nm gauge, leading to a perfectly scale-independent Newtonian bias.

%%%%%%%%%%%%%%%%%%%%%%%%%
\section{Conclusions}\label{sec:conclusions}

We have shown how the non-linear bias of Newtonian N-body simulations can be interpreted in terms of weak-field GR, being valid on all scales relevant for the formation of structures in the Universe. Our prescription allows the extraction of the bias in the Poisson gauge, which is a gauge that remains well defined in the weak-field limit of GR. 

In contrast to previous studies of bias in GR, we do not employ perturbation theory, but a weak-field description that remains accurate on smaller scales, where non-linear effects become important. 
When restricted to linear theory and to a universe only filled with matter, our results agree with the standard ones from the literature. In particular, within that limit,
we recover the well-known result that the simulation bias can be directly related to the relativistic bias in the synchronous-comoving gauge. 
By going beyond this simplified setup, by both including non-linear effects as well as the effects from relativistic species, we have shown that the GR dictionary has to be slightly updated, thus deviating from the standard results in the synchronous gauge. Apart from the employed Nm gauges, we also find that the Poisson gauge is a very suitable gauge for investigating the non-linear bias.

For the relativistic bias we detect a scale-dependency which kicks in on large scales, ranging from the scale of matter-radiation equality up to super-horizon scales. Essentially, this scale-dependency is the consequence of the matter collapse being not only governed by the matter overdensities but also affected by the metric perturbations.
In particular, the metric perturbations that are most relevant for the bias are the Bardeen potential $\Phi$, which deforms the local spatial volume element (w.r.t.\ to a Euclidean volume element), as well as the comoving curvature perturbation $\zeta$ which acts as a long wave-length perturbation for the collapse.

With the present methodology, the relativistic bias can be extracted from all standard Newtonian N-body simulations. In particular, Newtonian simulations can be interpreted 
in terms of weak-field GR within two orthogonal approaches,  called the forwards and backscaling approach. In the forward approach, the actual ICs for the Newtonian simulation are taken from the initialization redshift, whereas in the backscaling approach (which is very common in the literature) one initializes the Newtonian simulation by rescaling the linear final power spectrum back to the initialization redshift. Within our approach, we are able to provide a relativistic interpretation of the bias 
for both the forward and the backward approach. This is particularly simple in the backscaling approach as the complications are already present in the employed initial conditions and the simple dictionary  may be used (see section~\ref{sec:photons}).

Our weak-field approach allows not only the inclusion of non-linear effects but is also able to incorporate the effects of relativistic species.
We have already seen that the relativistic large-scale feature affects scales beyond the matter-radiation equality scale, linking it to the presence of radiation. But this feature is mediated via the metric potentials 'remembering' that the Universe underwent a phase of radiation domination before. In addition there is the direct impact that the remaining photon content has on a matter collapse. This effect was in fact computed in \cite{LoVerde:2014pxa} where a small but visible correction was found. Using the Nm gauge approach we confirm a feature on the largest scales, but we have shown that this particular feature is extremely small in the relativistic bias, and thus may safely be ignored. Massive neutrinos, by contrast, have very likely a non-negligible impact on the bias. We gave the first steps to incorporate the effects
of massive neutrinos, and argued that a backwards Nm approach applied to an actual Newtonian simulation with neutrinos should deliver an accurate description for the bias.

As a working example, we have applied our weak-field approach to Newtonian simulations for the matter collapse of a spherical-top hat perturbation. 
As expected, a purely Newtonian analysis does not yield to any scale-dependent features for the bias, whereas a GR description indeed reveals a
significant modification of the bias on large scales (see Figs.\,\ref{fig:NMG_bias} and~\ref{fig:NMG_bias_backwards}). By employing our weak-field description, we then recover permille-level agreement between the approaches on all considered scales.

Having obtained the framework to interpret the relativistic bias from Newtonian simulations, we plan to apply the techniques to actual N-body simulations (going beyond the spherical collapse). A particularly interesting topic would be to explore in depth the bias in the presence of massive neutrinos, and to investigate whether knowledge of the neutrino mass scale could be extracted from future large-scale structure surveys. Such avenues would require  high-resolution studies, 
obtained either from (relativistic) N-body simulations that actively include the evolution of massive neutrinos \cite{Adamek:2017uiq}, or from standard N-body simulations that incorporate the effects of massive neutrinos via the Nm-gauge framework \cite{Fidler:2018bkg}.

\section*{Acknowledgements}

We thank Julian Adamek for useful discussions, and the Technion Department of Physics for their hospitality and support.

\bibliographystyle{JHEP}
\bibliography{references}

\providecommand{\href}[2]{#2}\begingroup\raggedright\begin{thebibliography}{10}

\bibitem{Peacock:2000qk}
J.~A. Peacock and R.~E. Smith, \emph{{Halo occupation numbers and galaxy
  bias}}, \href{http://dx.doi.org/10.1046/j.1365-8711.2000.03779.x}{\emph{Mon.
  Not. Roy. Astron. Soc.} {\bf 318} (2000) 1144},
  [\href{http://arxiv.org/abs/astro-ph/0005010}{{\tt astro-ph/0005010}}].

\bibitem{Scoccimarro:2000gm}
R.~Scoccimarro, R.~K. Sheth, L.~Hui and B.~Jain, \emph{{How many galaxies fit
  in a halo? Constraints on galaxy formation efficiency from spatial
  clustering}}, \href{http://dx.doi.org/10.1086/318261}{\emph{Astrophys. J.}
  {\bf 546} (2001) 20--34}, [\href{http://arxiv.org/abs/astro-ph/0006319}{{\tt
  astro-ph/0006319}}].

\bibitem{Bernardeau:2001qr}
F.~Bernardeau, S.~Colombi, E.~Gaztanaga and R.~Scoccimarro, \emph{{Large scale
  structure of the universe and cosmological perturbation theory}},
  \href{http://dx.doi.org/10.1016/S0370-1573(02)00135-7}{\emph{Phys. Rept.}
  {\bf 367} (2002) 1--248}, [\href{http://arxiv.org/abs/astro-ph/0112551}{{\tt
  astro-ph/0112551}}].

\bibitem{Seljak:2004sj}
{\scshape SDSS} collaboration, U.~Seljak, A.~Makarov, R.~Mandelbaum, C.~M.
  Hirata, N.~Padmanabhan, P.~McDonald et~al., \emph{{SDSS galaxy bias from halo
  mass-bias relation and its cosmological implications}},
  \href{http://dx.doi.org/10.1103/PhysRevD.71.043511}{\emph{Phys. Rev.} {\bf
  D71} (2005) 043511}, [\href{http://arxiv.org/abs/astro-ph/0406594}{{\tt
  astro-ph/0406594}}].

\bibitem{Carbone:2008iz}
C.~Carbone, L.~Verde and S.~Matarrese, \emph{{Non-Gaussian halo bias and future
  galaxy surveys}}, \href{http://dx.doi.org/10.1086/592020}{\emph{Astrophys.
  J.} {\bf 684} (2008) L1--L4}, [\href{http://arxiv.org/abs/0806.1950}{{\tt
  0806.1950}}].

\bibitem{Matsubara:2011ck}
T.~Matsubara, \emph{{Nonlinear Perturbation Theory Integrated with Nonlocal
  Bias, Redshift-space Distortions, and Primordial Non-Gaussianity}},
  \href{http://dx.doi.org/10.1103/PhysRevD.83.083518}{\emph{Phys. Rev.} {\bf
  D83} (2011) 083518}, [\href{http://arxiv.org/abs/1102.4619}{{\tt
  1102.4619}}].

\bibitem{Desjacques:2016bnm}
V.~Desjacques, D.~Jeong and F.~Schmidt, \emph{{Large-Scale Galaxy Bias}},
  \href{http://dx.doi.org/10.1016/j.physrep.2017.12.002}{\emph{Phys. Rept.}
  {\bf 733} (2018) 1--193}, [\href{http://arxiv.org/abs/1611.09787}{{\tt
  1611.09787}}].

\bibitem{2009PhRvD..79l3507W}
D.~{Wands} and A.~{Slosar}, \emph{{Scale-dependent bias from primordial
  non-Gaussianity in general relativity}},
  \href{http://dx.doi.org/10.1103/PhysRevD.79.123507}{\emph{Phys. Rev.} {\bf
  D79} (June, 2009) 123507}, [\href{http://arxiv.org/abs/0902.1084}{{\tt
  0902.1084}}].

\bibitem{Yoo:2010ni}
J.~Yoo, \emph{{General Relativistic Description of the Observed Galaxy Power
  Spectrum: Do We Understand What We Measure?}},
  \href{http://dx.doi.org/10.1103/PhysRevD.82.083508}{\emph{Phys. Rev.} {\bf
  D82} (2010) 083508}, [\href{http://arxiv.org/abs/1009.3021}{{\tt
  1009.3021}}].

\bibitem{Baldauf:2011bh}
T.~Baldauf, U.~Seljak, L.~Senatore and M.~Zaldarriaga, \emph{{Galaxy Bias and
  non-Linear Structure Formation in General Relativity}},
  \href{http://dx.doi.org/10.1088/1475-7516/2011/10/031}{\emph{JCAP} {\bf 1110}
  (2011) 031}, [\href{http://arxiv.org/abs/1106.5507}{{\tt 1106.5507}}].

\bibitem{Bruni:2011ta}
M.~Bruni, R.~Crittenden, K.~Koyama, R.~Maartens, C.~Pitrou and D.~Wands,
  \emph{{Disentangling non-Gaussianity, bias and GR effects in the galaxy
  distribution}},
  \href{http://dx.doi.org/10.1103/PhysRevD.85.041301}{\emph{Phys. Rev.} {\bf
  D85} (2012) 041301}, [\href{http://arxiv.org/abs/1106.3999}{{\tt
  1106.3999}}].

\bibitem{2011PhRvD..84d3516C}
A.~{Challinor} and A.~{Lewis}, \emph{{Linear power spectrum of observed source
  number counts}},
  \href{http://dx.doi.org/10.1103/PhysRevD.84.043516}{\emph{{Phys. Rev.}} {\bf
  D84} (Aug., 2011) 043516}, [\href{http://arxiv.org/abs/1105.5292}{{\tt
  1105.5292}}].

\bibitem{Seljak:2008xr}
U.~Seljak, \emph{{Extracting primordial non-gaussianity without cosmic
  variance}},
  \href{http://dx.doi.org/10.1103/PhysRevLett.102.021302}{\emph{Phys. Rev.
  Lett.} {\bf 102} (2009) 021302}, [\href{http://arxiv.org/abs/0807.1770}{{\tt
  0807.1770}}].

\bibitem{Matarrese:2008nc}
S.~Matarrese and L.~Verde, \emph{{The effect of primordial non-Gaussianity on
  halo bias}}, \href{http://dx.doi.org/10.1086/587840}{\emph{Astrophys. J.}
  {\bf 677} (2008) L77--L80}, [\href{http://arxiv.org/abs/0801.4826}{{\tt
  0801.4826}}].

\bibitem{Adamek:2013wja}
J.~Adamek, D.~Daverio, R.~Durrer and M.~Kunz, \emph{{General Relativistic
  $N$-body simulations in the weak field limit}},
  \href{http://dx.doi.org/10.1103/PhysRevD.88.103527}{\emph{Phys. Rev.} {\bf
  D88} (2013) 103527}, [\href{http://arxiv.org/abs/1308.6524}{{\tt
  1308.6524}}].

\bibitem{Adamek:2016zes}
J.~Adamek, D.~Daverio, R.~Durrer and M.~Kunz, \emph{{gevolution: a cosmological
  N-body code based on General Relativity}},
  \href{http://dx.doi.org/10.1088/1475-7516/2016/07/053}{\emph{JCAP} {\bf 1607}
  (2016) 053}, [\href{http://arxiv.org/abs/1604.06065}{{\tt 1604.06065}}].

\bibitem{Adamek:2017uiq}
J.~Adamek, R.~Durrer and M.~Kunz, \emph{{Relativistic N-body simulations with
  massive neutrinos}},
  \href{http://dx.doi.org/10.1088/1475-7516/2017/11/004}{\emph{JCAP} {\bf 1711}
  (2017) 004}, [\href{http://arxiv.org/abs/1707.06938}{{\tt 1707.06938}}].

\bibitem{Brandbyge:2016raj}
J.~Brandbyge, C.~Rampf, T.~Tram, F.~Leclercq, C.~Fidler and S.~Hannestad,
  \emph{{Cosmological $N$-body simulations including radiation perturbations}},
  \href{http://dx.doi.org/10.1093/mnrasl/slw235}{\emph{Mon. Not. Roy. Astron.
  Soc.} {\bf 466} (2017) L68--L72},
  [\href{http://arxiv.org/abs/1610.04236}{{\tt 1610.04236}}].

\bibitem{Fidler:2015npa}
C.~Fidler, C.~Rampf, T.~Tram, R.~Crittenden, K.~Koyama and D.~Wands,
  \emph{{General relativistic corrections to $N$-body simulations and the
  Zel'dovich approximation}},
  \href{http://dx.doi.org/10.1103/PhysRevD.92.123517}{\emph{Phys. Rev.} {\bf
  D92} (2015) 123517}, [\href{http://arxiv.org/abs/1505.04756}{{\tt
  1505.04756}}].

\bibitem{Adamek:2017grt}
J.~Adamek, J.~Brandbyge, C.~Fidler, S.~Hannestad, C.~Rampf and T.~Tram,
  \emph{{The effect of early radiation in N-body simulations of cosmic
  structure formation}},
  \href{http://dx.doi.org/10.1093/mnras/stx1157}{\emph{Mon. Not. Roy. Astron.
  Soc.} (2017) }, [\href{http://arxiv.org/abs/1703.08585}{{\tt 1703.08585}}].

\bibitem{Fidler:2016tir}
C.~Fidler, T.~Tram, C.~Rampf, R.~Crittenden, K.~Koyama and D.~Wands,
  \emph{{Relativistic Interpretation of Newtonian Simulations for Cosmic
  Structure Formation}},
  \href{http://dx.doi.org/10.1088/1475-7516/2016/09/031}{\emph{JCAP} {\bf 1609}
  (2016) 031}, [\href{http://arxiv.org/abs/1606.05588}{{\tt 1606.05588}}].

\bibitem{Fidler:2017ebh}
C.~Fidler, T.~Tram, C.~Rampf, R.~Crittenden, K.~Koyama and D.~Wands,
  \emph{{Relativistic initial conditions for N-body simulations}},
  \href{http://dx.doi.org/10.1088/1475-7516/2017/06/043}{\emph{JCAP} {\bf 1706}
  (2017) 043}, [\href{http://arxiv.org/abs/1702.03221}{{\tt 1702.03221}}].

\bibitem{Fidler:2017pnb}
C.~Fidler, T.~Tram, C.~Rampf, R.~Crittenden, K.~Koyama and D.~Wands,
  \emph{{General relativistic weak-field limit and Newtonian N-body
  simulations}},
  \href{http://dx.doi.org/10.1088/1475-7516/2017/12/022}{\emph{JCAP} {\bf 1712}
  (2017) 022}, [\href{http://arxiv.org/abs/1708.07769}{{\tt 1708.07769}}].

\bibitem{Fidler:2018bkg}
C.~Fidler, A.~Kleinjohann, T.~Tram, C.~Rampf and K.~Koyama, \emph{{A new
  approach to cosmological structure formation with massive neutrinos}},
  \href{http://arxiv.org/abs/1807.03701}{{\tt 1807.03701}}.

\bibitem{Kodama:1985bj}
H.~Kodama and M.~Sasaki, \emph{{Cosmological Perturbation Theory}},
  \href{http://dx.doi.org/10.1143/PTPS.78.1}{\emph{Prog. Theor. Phys. Suppl.}
  {\bf 78} (1984) 1--166}.

\bibitem{Malik:2008im}
K.~A. Malik and D.~Wands, \emph{{Cosmological perturbations}},
  \href{http://dx.doi.org/10.1016/j.physrep.2009.03.001}{\emph{Phys. Rept.}
  {\bf 475} (2009) 1--51}, [\href{http://arxiv.org/abs/0809.4944}{{\tt
  0809.4944}}].

\bibitem{Villa:2015ppa}
E.~Villa and C.~Rampf, \emph{{Relativistic perturbations in $\Lambda$CDM:
  Eulerian \& Lagrangian approaches}},
  \href{http://dx.doi.org/10.1088/1475-7516/2016/01/030}{\emph{JCAP} {\bf 1601}
  (2016) 030}, [\href{http://arxiv.org/abs/1505.04782}{{\tt 1505.04782}}].

\bibitem{Yoo:2014sfa}
J.~Yoo and M.~Zaldarriaga, \emph{{Beyond the Linear-Order Relativistic Effect
  in Galaxy Clustering: Second-Order Gauge-Invariant Formalism}},
  \href{http://dx.doi.org/10.1103/PhysRevD.90.023513}{\emph{Phys. Rev.} {\bf
  D90} (2014) 023513}, [\href{http://arxiv.org/abs/1406.4140}{{\tt
  1406.4140}}].

\bibitem{Bertacca:2014dra}
D.~Bertacca, R.~Maartens and C.~Clarkson, \emph{{Observed galaxy number counts
  on the lightcone up to second order: I. Main result}},
  \href{http://dx.doi.org/10.1088/1475-7516/2014/09/037}{\emph{JCAP} {\bf 1409}
  (2014) 037}, [\href{http://arxiv.org/abs/1405.4403}{{\tt 1405.4403}}].

\bibitem{Yoo:2014vta}
J.~Yoo, \emph{{Proper-time hypersurface of nonrelativistic matter flows: Galaxy
  bias in general relativity}},
  \href{http://dx.doi.org/10.1103/PhysRevD.90.123507}{\emph{Phys. Rev.} {\bf
  D90} (2014) 123507}, [\href{http://arxiv.org/abs/1408.5137}{{\tt
  1408.5137}}].

\bibitem{Rampf:2014mga}
C.~Rampf and A.~Wiegand, \emph{{Relativistic Lagrangian displacement field and
  tensor perturbations}},
  \href{http://dx.doi.org/10.1103/PhysRevD.90.123503}{\emph{Phys. Rev.} {\bf
  D90} (2014) 123503}, [\href{http://arxiv.org/abs/1409.2688}{{\tt
  1409.2688}}].

\bibitem{Bertacca:2015mca}
D.~Bertacca, N.~Bartolo, M.~Bruni, K.~Koyama, R.~Maartens, S.~Matarrese et~al.,
  \emph{{Galaxy bias and gauges at second order in General Relativity}},
  \href{http://dx.doi.org/10.1088/0264-9381/32/17/175019}{\emph{Class. Quant.
  Grav.} {\bf 32} (2015) 175019}, [\href{http://arxiv.org/abs/1501.03163}{{\tt
  1501.03163}}].

\bibitem{Rampf:2016wom}
C.~Rampf, E.~Villa, D.~Bertacca and M.~Bruni, \emph{{Lagrangian theory for
  cosmic structure formation with vorticity: Newtonian and post-Friedmann
  approximations}},
  \href{http://dx.doi.org/10.1103/PhysRevD.94.083515}{\emph{Phys. Rev.} {\bf
  D94} (2016) 083515}, [\href{http://arxiv.org/abs/1607.05226}{{\tt
  1607.05226}}].

\bibitem{Brustein:2011dy}
R.~Brustein and A.~Riotto, \emph{{Evolution Equation for Non-linear
  Cosmological Perturbations}},
  \href{http://dx.doi.org/10.1088/1475-7516/2011/11/006}{\emph{JCAP} {\bf 1111}
  (2011) 006}, [\href{http://arxiv.org/abs/1105.4411}{{\tt 1105.4411}}].

\bibitem{Kopp:2013tqa}
M.~Kopp, C.~Uhlemann and T.~Haugg, \emph{{Newton to Einstein -- dust to dust}},
  \href{http://dx.doi.org/10.1088/1475-7516/2014/03/018}{\emph{JCAP} {\bf 1403}
  (2014) 018}, [\href{http://arxiv.org/abs/1312.3638}{{\tt 1312.3638}}].

\bibitem{Goldberg:2017gsm}
S.~R. Goldberg, C.~Gallagher and T.~Clifton, \emph{{Perturbation theory for
  cosmologies with non-linear structure}},
  \href{http://arxiv.org/abs/1707.01042}{{\tt 1707.01042}}.

\bibitem{Goldberg:2016lcq}
S.~R. Goldberg, T.~Clifton and K.~A. Malik, \emph{{Cosmology on all scales: a
  two-parameter perturbation expansion}},
  \href{http://dx.doi.org/10.1103/PhysRevD.95.043503}{\emph{Phys. Rev.} {\bf
  D95} (2017) 043503}, [\href{http://arxiv.org/abs/1610.08882}{{\tt
  1610.08882}}].

\bibitem{Mukhanov:1990me}
V.~F. Mukhanov, H.~A. Feldman and R.~H. Brandenberger, \emph{{Theory of
  cosmological perturbations. Part 1. Classical perturbations. Part 2. Quantum
  theory of perturbations. Part 3. Extensions}},
  \href{http://dx.doi.org/10.1016/0370-1573(92)90044-Z}{\emph{Phys. Rept.} {\bf
  215} (1992) 203--333}.

\bibitem{Ma:1995ey}
C.-P. Ma and E.~Bertschinger, \emph{{Cosmological perturbation theory in the
  synchronous and conformal Newtonian gauges}},
  \href{http://dx.doi.org/10.1086/176550}{\emph{Astrophys. J.} {\bf 455} (1995)
  7--25}, [\href{http://arxiv.org/abs/astro-ph/9506072}{{\tt
  astro-ph/9506072}}].

\bibitem{Chisari:2011iq}
N.~E. Chisari and M.~Zaldarriaga, \emph{{Connection between Newtonian
  simulations and general relativity}},
  \href{http://dx.doi.org/10.1103/PhysRevD.84.089901,
  10.1103/PhysRevD.83.123505}{\emph{Phys. Rev.} {\bf D83} (2011) 123505},
  [\href{http://arxiv.org/abs/1101.3555}{{\tt 1101.3555}}].

\bibitem{LoVerde:2014pxa}
M.~LoVerde, \emph{{Halo bias in mixed dark matter cosmologies}},
  \href{http://dx.doi.org/10.1103/PhysRevD.90.083530}{\emph{Phys. Rev.} {\bf
  D90} (2014) 083530}, [\href{http://arxiv.org/abs/1405.4855}{{\tt
  1405.4855}}].

\end{thebibliography}\endgroup

\end{document}